\documentclass[twocolumn,showpacs,preprintnumbers,amsmath,amssymb,superscriptaddress]{revtex4}

\usepackage{graphicx}
\usepackage{bm}

\def\mib#1{\hbox{\boldmath $#1$}}

\def\eq#1{Eq.\,(\ref{#1})}

\begin{document}

\preprint{APS/123-QED}

\title{ Faddeev calculation
of \mib{\hbox{}^{\ 6}_{\Lambda \Lambda} \hbox{He}}
using \mib{SU_6} quark-model baryon-baryon interactions}

\author{Y. Fujiwara}
\affiliation{Department of Physics, Kyoto University, 
Kyoto 606-8502, Japan}%
 \email{fujiwara@ruby.scphys.kyoto-u.ac.jp}

\author{M. Kohno}%
\affiliation{Physics Division, Kyushu Dental College,
Kitakyushu 803-8580, Japan}

\author{K. Miyagawa}%
\affiliation{Department of Applied Physics,
Okayama Science University, Okayama 700-0005, Japan}

\author{Y. Suzuki}%
\affiliation{Department of Physics, Niigata University,
Niigata 950-2181, Japan}

\author{J.-M. Sparenberg}%
\affiliation{TRIUMF, 4004 Wesbrook Mall,
Vancouver, British Columbia, Canada V6T 2A3}

\date{\today}

\begin{abstract}
Quark-model hyperon-nucleon and hyperon-hyperon interactions
by the Kyoto-Niigata group are applied
to the two-$\Lambda$ plus $\alpha$ system
in a new three-cluster Faddeev formalism using two-cluster
resonating-group method kernels.
The model fss2 gives a reasonable two-$\Lambda$ separation
energy $\Delta B_{\Lambda \Lambda}=1.41$ MeV, which is consistent
with the recent empirical value,
$\Delta B^{\rm exp}_{\Lambda \Lambda}
=1.01 \pm 0.20$ MeV, deduced from the Nagara event.
Some important effects that are not taken into account
in the present calculation are discussed.
\end{abstract}

\maketitle



A new discovery of the double $\Lambda$ hypernuclei,
$\hbox{}^{\ 6}_{\Lambda \Lambda} \hbox{He}$,
called the Nagara event \cite{TA01} has provided
an invaluable source of information for the strength
of the $\Lambda \Lambda$ interaction. Before this discovery,
it had been believed that the two-$\Lambda$ separation
energy measured by $\Delta B_{\Lambda \Lambda}
=B_{\Lambda \Lambda}(\hbox{}^{\ \,6}_{\Lambda \Lambda}\hbox{He})
-2B_{\Lambda}(\hbox{}^{\,5}_{\Lambda}\hbox{He})$ or
equivalently by $2E(\hbox{}^5\hbox{He})
-E(\hbox{}^{\ \,6}_{\Lambda \Lambda}\hbox{He})
-E(\hbox{}^4\hbox{He})$ was fairly large,
$\Delta B_{\Lambda \Lambda} \sim 4.3$ MeV, which implies
that the $\Lambda \Lambda$ interaction is more attractive
than the corresponding $\hbox{}^1S_0$ $\Lambda N$ interaction.
It was argued in Ref.\ \cite{CA97} that the proper treatment
of the $\Lambda \Lambda$-$\Xi N$ coupling effect
in the $\Lambda \Lambda \alpha$ model is important to
reproduce this $\Delta B_{\Lambda \Lambda}$ value in the
coupled-channel AGS formalism using the $\hbox{}^1S_0$ 
$\Lambda \Lambda$ interaction of the Nijmegen model D.  
Now it is clear that the Nijmegen model D is not appropriate
to describe the double $\Lambda$ hypernuclei.
Almost unique identification of the sequential decay
processes involved in the Nagara event
enforced it necessary to reanalyze the
previous three events of the double $\Lambda$ hypernuclei
\cite{PR66,DA63,AO91} and led to the conclusion that
the $\Lambda \Lambda$ interaction is actually weakly attractive,
under the assumption of possible involvement of excited states
in the intermediate processes.
The $\Delta B_{\Lambda \Lambda}$ value deduced from
the Nagara event is $1.01 \pm 0.20$ MeV \cite{TA01}.

Based on this experimental information, several calculations
have been carried out to determine the strength
of the $\Lambda \Lambda$ interaction precisely
and to find an appropriate interaction model
mainly among the meson-theoretical
Nijmegen models. For example, Filikhin, Gal,
and Suslov \cite{FI03} performed detailed Faddeev calculations
using the $\Lambda \Lambda \alpha$ cluster model
with many phenomenological $\Lambda \Lambda$ interactions
and the so-called Isle $\Lambda \alpha$ potential with
a repulsion core.
They used the $S$-wave $\Lambda \alpha$ and
$\Lambda \Lambda$ potentials for all the allowed partial waves.
Since $\hbox{}^{\ 6}_{\Lambda \Lambda} \hbox{He}$ is essentially
an $S$-wave dominant system, their approximation is legitimate.   
Nevertheless, the Nijmegen soft-core model
NSC97e \cite{NSC97} was found to have too weak $\Lambda \Lambda$
interaction, corresponding
to $\Delta B_{\Lambda \Lambda} \sim 0.66$ MeV \cite{FI03}.

We have discussed in Ref.\ \cite{rearr} that the cluster model
calculation with the $\alpha$ cluster needs a special care
with an important rearrangement effect originating mainly
from the starting energy dependence of the $G$-matrix interaction,
when we consider composite-particle interactions starting from
bare baryon-baryon interactions.
For example, the energy loss of the interaction term
in $\hbox{}^5\hbox{He}$ due to the added $\Lambda$ particle
is estimated to be 2.5 - 2.9 MeV in the model-independent way.
This effect plays a major role to explain the well-known
overbinding phenomena of the $\hbox{}^5\hbox{He}$. 
This effect is renormalized in usual $\Lambda \alpha$ potentials
by fitting the $\Lambda$ separation
energy $B_\Lambda(\hbox{}^5_\Lambda \hbox{He})
=3.12\pm 0.02~\hbox{MeV}$.
In the $\Lambda \Lambda \alpha$ system,
however, there still remains an unrenormalizable effect mainly
originating from the starting energy dependence
of the $\Lambda N$ interaction, which is found to be
a repulsive effect of about 1 MeV \cite{rearr}.
As the result, the $S$-state matrix element
of the $\Lambda \Lambda$ interaction is
not $-\Delta B_{\Lambda \Lambda} \sim -1~\hbox{MeV}$, but
should be larger than $-2$ MeV. From this argument, we can
conclude that the $\hbox{}^1S_0$ $\Lambda \Lambda$ interaction
of NSC97e is by far too weak, and there is no meson-theoretical
models available to explain the Nagara event.

The purpose of this brief report is to show the extent
how our quark-model baryon-baryon interaction
fss2 \cite{fss2,B8B8} can give a consistent description
of the $\Lambda N$ and $\Lambda \Lambda$ interactions
with the available experimental data of light
single- and double-$\Lambda$ hypernuclei.
The model fss2 describes all the available
nucleon-nucleon ($NN$) and hyperon-nucleon ($YN$) scattering data,
by incorporating the effective meson-exchange potentials
at the quark level. It is now extended to the arbitrary
two-baryon systems of the octet baryons without introducing
any extra parameters \cite{B8B8}.
The strangeness $S=-2$ sector, in particular, involves
several important aspects of the baryon-baryon interactions.
First it contains the $\Lambda \Lambda$ interaction,
whose knowledge is essential to understand the binding
mechanism of the double-$\Lambda$ hypernuclei.
The second is that the isospin $T=0$ system corresponds to the
so-called $H$-particle channel, in which a strong
attraction is expected from the color-magnetic interaction
of the quark model. The third is the existence of the
Pauli-forbidden state at the quark level,
with the $SU_3$ quantum number $(11)_s$. The existence of
such Pauli-forbidden state usually implies a strong
repulsion in some particular channels.
It is therefore important to deal with the effect of the
Pauli principle properly in the quark-model
baryon-baryon interactions.
Here we carry out Faddeev calculations
of the $\Lambda \Lambda \alpha$ system,
by directly using the quark-model baryon-baryon interactions
in the strangeness $S=-2$ sector, and show that
the $\Lambda \Lambda$ interaction
of fss2 is consistent with the Nagara event after several
corrections which are not easily incorporated in the
present calculation.

The three-cluster Faddeev formalism used here is
recently developed for general three-cluster systems
interacting via two-cluster resonating-group
method (RGM) kernels \cite{TRGM,RED}.
A nice point of this formalism is that the
underlying $NN$, $YN$, and hyperon-hyperon ($YY$) interactions
are more directly related to the structure of the hypernuclei
than the models assuming simple two-cluster potentials.
The reliability of this formalism is already confirmed
in several systems; i.e., the three-nucleon
bound state \cite{triton}, the hypertriton \cite{hypt},
the $3 \alpha$ and $\Lambda \alpha \alpha$ systems \cite{BE9L}.
The last application involves an effective $\Lambda N$ force,
called the SB force, which is a simple two-range Gaussian
potential generated from the phase-shift behavior of fss2,
by using an inversion method based on supersymmetric
quantum mechanics \cite{SB97}.
It is given by
\begin{eqnarray}
v(\hbox{}^1E) & = & -128.0~\exp(-0.8908~r^2)\nonumber \\
& & +1015~\exp(-5.383~r^2)\ ,\nonumber \\
v(\hbox{}^3E) & = & -56.31~f~\exp(-0.7517~r^2)\nonumber \\
& & +1072~\exp(-13.74~r^2)\ ,
\label{eq1}
\end{eqnarray}
where $r$ is the relative distance
between $\Lambda$ and $N$ in fm and the energy is measured
in MeV. The odd interaction is assumed to be zero (pure Serber type).
We generate the $\Lambda \alpha$ potential by folding these
with the simple $(0s)^4$ shell-model wave function
of the $\alpha$ cluster.
In \eq{eq1} an adjustable parameter $f$ is introduced
to circumvent the overbinding problem of $\hbox{}^5_\Lambda \hbox{He}$.
The value $f=0.8923$ is necessary to reproduce the empirical
value $B_\Lambda(\hbox{}^5_\Lambda \hbox{He})=3.120~\hbox{MeV}$,
when the harmonic oscillator width parameter
of the $\alpha$ cluster is assumed
to be $\nu=0.257~\hbox{fm}^{-2}$.
By using this $\Lambda N$ force and the $\alpha \alpha$ RGM
kernel generated from the three-range Minnesota force,
we have shown in Ref.\ \cite{BE9L} that
the the mutually related, $\alpha \alpha$,
$3\alpha$, and $\alpha \alpha \Lambda$ systems are well
reproduced in terms of a unique set of the baryon-baryon
interactions. In particular, the ground-state and excitation
energies of $\hbox{}^9_\Lambda \hbox{Be}$ are
reproduced within 100 $\sim 200$ keV accuracy.

The total wave function of the $\Lambda \Lambda \alpha$ system
is expressed as the superposition of two independent Faddeev
components $\psi$ and $\varphi$:
$\Psi=\psi+(1-P_{12})\varphi$. The two $\Lambda$ particles are
numbered 1 and 2, the $\alpha$-cluster is numbered 3.
The Faddeev equation reads 
\begin{eqnarray}
\psi & = & G_0 \widetilde{T}_{\Lambda \Lambda}
(\varepsilon_{\Lambda \Lambda})
(1-P_{12}) \varphi\ ,\nonumber \\
\varphi & = & G_0 T_{\Lambda \alpha} \left(\psi 
-P_{12} \varphi \right)\ .
\label{eq2}
\end{eqnarray}
Here, $\widetilde{T}_{\Lambda \Lambda}
(\varepsilon_{\Lambda \Lambda})$ is
the $\Lambda \Lambda$ component of the redundancy-free
$\Lambda \Lambda$-$\Xi N$-$\Sigma \Sigma$ $\widetilde{T}$-matrices
in the specific channel with the strangeness $S=-2$ and
the isospin $T=0$. These $T$-matrices are generated
from the RGM kernel of the $YY$ interaction,
$V^{\rm RGM}_{YY}(\varepsilon_{YY})$, by solving the full
coupled-channel Lippmann-Schwinger equation in the momentum space.
The elimination of the Pauli-forbidden state with the $SU_3$ quantum
number $(11)_s$ is automatically taken care of, simply by using
the ``RGM'' $T$-matrix, $\widetilde{T}_{\Lambda \Lambda}
(\varepsilon_{\Lambda \Lambda})$ according
to the prescription given in Ref.\ \cite{TRGM}.
The total wave function $\Psi$ is orthogonal to this
Pauli-forbidden state, if we formulate a full coupled-channel
Faddeev equation for the $\Lambda \Lambda
\alpha$-$\Xi N \alpha$-$\Sigma \Sigma \alpha$ system.
Such a calculation is not feasible for the time being,
since we also need the $N \alpha$, $\Xi \alpha$,
and $\Sigma \alpha$ interactions.
Here we simply use the $\Lambda \Lambda$ component of
the redundancy-free $\widetilde{T}$-matrix.
The energy dependence involved in the RGM kernel and
the $\widetilde{T}$-matrix is treated
self-consistently by calculating the matrix elements
of the (quark-model) $\Lambda \Lambda$ Hamiltonian as
\begin{equation}
\varepsilon_{\Lambda \Lambda}=\langle \Psi|h_{\Lambda \Lambda}
+V^{\rm RGM}_{\Lambda \Lambda}(\varepsilon_{\Lambda \Lambda})
|\Psi \rangle \ \ .
\label{eq3}
\end{equation}
The detailed prescription for the energy dependence of the
RGM kernel and the Pauli-forbidden state in the quark-model
baryon-baryon interaction is given in Ref.\ \cite{hypt}.
A Faddeev formalism involving two identical
particles (or clusters) are spelled out in Ref.\ \cite{BE9L}.

Since we are interested in the $J^\pi=0^+$ ground state
with the isospin $T=0$, the channel specification scheme
of the $\Lambda \Lambda \alpha$ system is very simple.
It becomes even simpler if we introduce no non-central forces
since the $\Lambda \alpha$ interaction is known to involve 
a very weak spin-orbit force.
In the $(\Lambda \Lambda)$-$\alpha$ channel,
the exchange symmetry of the two $\Lambda$'s
requires $(-)^{\lambda+S}=1$, where $\lambda$ and $S$ are
the relative orbital angular-momentum and spin values
of the two-$\Lambda$ subsystem.
The possible two-$\lambda$ states are therefore
$\hbox{}^1\lambda_\lambda$ ($\lambda$=even) for $S=0$ and
$\hbox{}^3\lambda_\lambda$ ($\lambda$=odd) for $S=1$.
If we neglect non-central forces, the spin value $S$
and the total orbital angular-momentum quantum
number $L$ are good quantum numbers,
and only $\hbox{}^1S_0$, $\hbox{}^1D_2$,
$\hbox{}^1G_4$, $\cdots$ states
of the $\Lambda \Lambda$ interaction contribute
in the ground state with $L=S=0$.
Note that the orbital angular-momentum of the
$\alpha$ particle, $\ell$, is equal to $\lambda$ since $J=0$.
Similarly, in the $(\Lambda \alpha)$-$\Lambda$ channel,
the relative angular-momentum of the $\Lambda \alpha$ subsystem,
$\ell_1$, is equal to the orbital
angular momentum of the spectator $\Lambda$, $\ell_2$,
because of the parity conservation and the possible
spin value, $S=0$ or 1.
These simplifications are of course the result of the channel
truncation that we do not include the coupling
to the possible $\Xi N \alpha$
and $\Sigma \Sigma \alpha$ configurations,
in the present $\Lambda \Lambda \alpha$ model space.
All the partial waves up
to $\lambda_{\rm max}$-${\ell_1}_{\rm max}$=6-6 are included
for $\lambda=\ell$ and $\ell_1=\ell_2$.
The momentum discretization points with $n_1$-$n_2$-$n_3$=10-10-5
in the previous notation \cite{BE9L} are used for solving
the Faddeev equation. This ensures 1 keV accuracy.

\begin{table}[t]
\caption{Comparison of $\Delta B_{\Lambda \Lambda}$ values
in MeV, predicted by various $\Lambda \Lambda$ interactions
and $V_{\Lambda N}$ potentials.
The $\Lambda \Lambda$ potential $V_{\Lambda \Lambda}$(Hiyama) is
the three-range Gaussian potential
used in Ref.\ \protect\cite{HI97},
and $V_{\Lambda \Lambda}$(SB) the two-range Gaussian potential
given in \protect\eq{eq4}.
FSS and fss2 use the $\Lambda \Lambda$ RGM $T$-matrix
in the free space, with $\varepsilon_{\Lambda \Lambda}$ being 
the $\Lambda \Lambda$ expectation value determined self-consistently.
$\Delta B^{\rm exp}_{\Lambda \Lambda}
=1.01 \pm 0.20$ MeV \protect\cite{TA01}.
}
\renewcommand{\arraystretch}{1.2}
\setlength{\tabcolsep}{2mm}
\begin{center}
\begin{tabular}{ccccccc}
\hline
$V_{\Lambda \Lambda}$ & Hiyama
& \multicolumn{2}{c}{FSS} & \multicolumn{2}{c}{fss2}
& SB \\
\hline
$V_{\Lambda N}$ & $\Delta B_{\Lambda \Lambda}$
 & $\Delta B_{\Lambda \Lambda}$ & $\varepsilon_{\Lambda \Lambda}$
 & $\Delta B_{\Lambda \Lambda}$ & $\varepsilon_{\Lambda \Lambda}$
 & $\Delta B_{\Lambda \Lambda}$ \\
\hline
SB & 3.618 &  3.657 & 5.124 & 1.413 & 5.938 & 1.910 \\
NS & 3.548 &  3.630 & 5.151 & 1.366 & 5.947 & 1.914 \\
ND & 3.181 &  3.237 & 4.479 & 1.288 & 5.229 & 1.645 \\
NF & 3.208 &  3.305 & 4.622 & 1.271 & 5.407 & 1.713 \\
JA & 3.370 &  3.473 & 4.901 & 1.307 & 5.702 & 1.824 \\
JB & 3.486 &  3.599 & 5.141 & 1.327 & 5.952 & 1.911 \\
\hline
\end{tabular}
\label{table1}
\end{center}
\end{table}

\begin{figure}[t]
\begin{minipage}[t]{85mm}
\includegraphics[angle=-90,width=82mm]{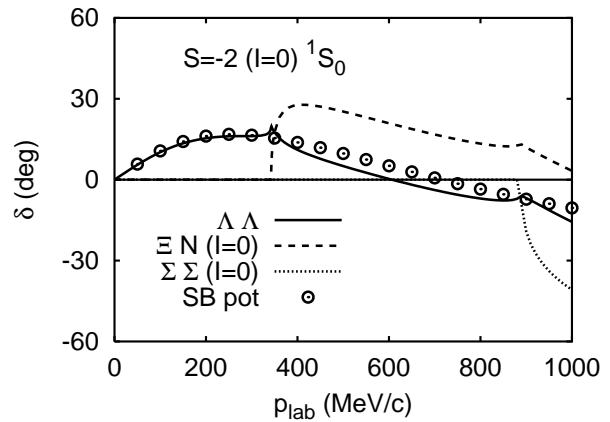}
\caption{
$\hbox{}^1S_0$ phase shifts, predicted by fss2,
in the $\Lambda \Lambda$-$\Xi N$-$\Sigma \Sigma$ coupled-channel
system with the isospin $I=0$.
The single-channel phase shift of the $\Lambda \Lambda$ scattering,
predicted by the SB potential, is also shown in circles.}
\label{fig1}
\end{minipage}
\end{figure}

Table \ref{table1} shows the $\Delta B_{\Lambda \Lambda}$ values
in MeV, predicted by various combinations
of the $\Lambda N$ and $\Lambda \Lambda$ interactions.
The results of a simple three-range Gaussian potential,
$V_{\Lambda \Lambda}$(Hiyama), used in Ref.\ \cite{HI97}
are also shown.
We find that this $\Lambda \Lambda$ potential and
the RGM $T$-matrix for the old version of our
quark-model interaction FSS \cite{FSS} yield
very similar results with the
large $\Delta B_{\Lambda \Lambda}$ values about 3.6 MeV,
since the $\Lambda \Lambda$ phases shifts
predicted by these interactions
increase up to about $40^\circ$. The improved quark model fss2
yields $\Delta B_{\Lambda \Lambda}=1.41$ MeV.
The energy gain due to the expansion of the partial waves
from the $S$-wave to the $I$-wave is $35 \sim -50$ keV,
depending on the weakly attractive or repulsive nature
of the $P$-wave $\Lambda N$ force. 
In Table \ref{table1}, results are also shown
for $V_{\Lambda \Lambda}$(SB), which is a two-range
Gaussian potential generated
from the $\hbox{}^1S_0$ $\Lambda \Lambda$ phase
shift of fss2, by using the supersymmetric
inversion method \cite{SB97}.
This potential is given by
\begin{eqnarray}
V_{\Lambda \Lambda}({\rm SB}) & = & -103.9~\exp(-1.176~r^2)
\nonumber \\
& & +658.2~\exp(-5.936~r^2)\ ,
\label{eq4}
\end{eqnarray}
where $r$ is the relative distance
between two $\Lambda$'s in fm and the energy in MeV. 
This potential reproduces the low-energy behavior
of the $\Lambda \Lambda$ phase shift of fss2 quite well,
as seen in Fig.\ \ref{fig1}.
We use this for all even partial waves and set the odd
components zero by assuming the pure Serber type.
[The odd components of the $\Lambda \Lambda$ interaction
give no contribution to the present calculation
in the $LS$ coupling scheme anyway.]
We find that this $\Lambda \Lambda$ potential yields
larger $\Delta B_{\Lambda \Lambda}$ values than the
fss2 RGM $T$-matrix by 0.36 MeV - 0.58 MeV.
We think that this difference of around 0.5 MeV between
our fss2 result and the $V_{\Lambda \Lambda}$(SB) result
is probably because
we neglected the full coupled-channel effects
of the $\Lambda \Lambda \alpha$ channel
to the $\Xi N \alpha$ and $\Sigma \Sigma \alpha$ channels.
In our previous Faddeev calculation
for $\hbox{}^3\hbox{H}$ \cite{triton}, the energy gain
due to the increase of the partial waves
from the 2-channel ($S$-wave only) to
5-channel ($S+D$ waves) calculations
is 0.36 - 0.38 MeV (see Table III).
We should keep in mind that in all of these
three-cluster calculations
the Brueckner rearrangement effect of the $\alpha$-cluster
with the magnitude of about 1 MeV (repulsive)
is very important \cite{rearr}.
It is also reported in Ref.\ \cite{SU99} that
the quark Pauli effect between the $\alpha$ cluster
and the $\Lambda$ hyperon yields a non-negligible
repulsive contribution of 0.1 - 0.2 MeV for
the $\Lambda$ separation energy
of $\hbox{}^{\ 6}_{\Lambda \Lambda} \hbox{He}$,
even when a rather compact $(3q)$ size of $b \sim 0.6$ fm
is assumed as in our quark-model interactions.
Taking all of these effects into consideration, we can conclude
that the present results by fss2 are in good agreement
with the experimental value,
$\Delta B^{\rm exp}_{\Lambda \Lambda}
=1.01 \pm 0.20$ MeV, by the Nagara event \cite{TA01}.

Table \ref{table2} lists the energy decomposition to kinetic-
and potential-energy contributions for the SB $\Lambda N$ force.
We find that the $\Lambda \Lambda$ potential matrix element
in fss2 is $-2.4 \sim -2.6$ MeV, which is much smaller then
that of FSS and the Hiyama potential ($-6 \sim -7$ MeV).
This is consistent with the $(0s)$ matrix element
of the $\Lambda \Lambda$ $G$-matrix of fss2 \cite{rearr},
$\langle (0s)^2|G_{\Lambda \Lambda}|(0s)^2 \rangle
=-2.95~\hbox{MeV}$, obtained for the free-space $G$-matrix
calculation with $\nu=0.25~\hbox{fm}^{-2}$.
For the normal nucleon density, $\rho_0=1.35~\hbox{fm}^{-1}$,
this value is slightly reduced to $-2.83~\hbox{MeV}$.
If $\rho_0$ is also assumed for $\Lambda$,
it is further reduced to $-2.63~\hbox{MeV}$.
If we compare the $\Lambda \alpha$ kinetic-energy matrix
elements in fss2 (8.553 MeV) and
in $V_{\Lambda \Lambda}$(SB) (8.774 MeV) with that
of the $E(\hbox{}^{9}_{\Lambda}\hbox{Be})$ system
in Ref.\ \cite{BE9L}, 9.215 MeV [see Eq.\,(39)],
the latter is a little larger since the $\Lambda$ is more strongly
attracted by the two $\alpha$ clusters.
The $\varepsilon_{\Lambda \alpha}$ value $-2.5 \sim -2.8$ MeV
in Table \ref{table2} should
be compared with the free value $-3.12$ MeV
in $\hbox{}^5_{\Lambda}\hbox{He}$, but the decomposition to the
kinetic-energy and potential-energy contributions is fairly
different since $E_{\Lambda \alpha}=3.854-6.974
=-3.120~\hbox{MeV}$ in $\hbox{}^5_{\Lambda}\hbox{He}$.

The $G$-matrix calculation of fss2 also shows that
the channel coupling effect to the $\Lambda \Lambda$ matrix element
is 0.5 $\sim$ 1 MeV, and the Pauli blocking effect
of the $\Xi N$ channel is about 0.2 MeV.
The latter is almost half of the 0.43 MeV,
claimed in Ref.\ \cite{MY03}.
We can also carry out the Faddeev calculation by
switching off the channel coupling in the $T$-matrix
calculations. The results for the fss2 $\Lambda \Lambda$
and SB $\Lambda N$ model are $\Delta B_{\Lambda \Lambda}
=1.141$ MeV for the $\Lambda \Lambda$ single-channel
calculation and 1.454 MeV for
the $\Lambda \Lambda$-$\Xi N$ double-channel calculation.
The energy gain by the full coupled-channel $T$-matrix
calculation is only 0.27 MeV.
However, such truncation of channels spoils the
exact treatment of the Pauli principle, and
the RGM $T$-matrix does not satisfy
the orthogonality condition to the
Pauli forbidden $(11)_s$ state.

\begin{table}[t]
\caption{Decomposition of the ground-state energy
of $\hbox{}^{\ \,6}_{\Lambda \Lambda}\hbox{He}$ ($E$),
and the $\Lambda \Lambda$ ($\varepsilon_{2\Lambda})$ and
$\Lambda \alpha$ ($\varepsilon_{\Lambda \alpha}$) expectation
values to the the kinetic- and potential-energy contributions.
The SB $\Lambda N$ force is used.
The unit is in MeV. The experimental value is
$E^{\rm exp}=-7.25 \pm 0.19$ MeV \protect\cite{TA01}.
}
\label{table2}
\begin{center}
\renewcommand{\arraystretch}{1.2}
\setlength{\tabcolsep}{5mm}
\begin{tabular}{cl}
\hline
$V_{\Lambda \Lambda}$ & kinetic $+$ potential $=$ total \\
\hline
       & $E=21.328-31.186=-9.858$ \\
Hiyama & $\varepsilon_{2\Lambda}=11.869-6.779=5.089$ \\
       & $\varepsilon_{\Lambda \alpha}=10.388-12.203=-1.815$ \\
\hline
       & $E=20.179-30.076=-9.897$ \\
FSS    & $\varepsilon_{2\Lambda}=10.800-5.676=5.124$ \\
       & $\varepsilon_{\Lambda \alpha}=9.927-12.200=-2.274$ \\
\hline
       & $E=17.111-24.764=-7.653$ \\
fss2   & $\varepsilon_{2\Lambda}=8.567-2.628=5.938$ \\
       & $\varepsilon_{\Lambda \alpha}=8.553-11.068=-2.515$ \\
\hline
       & $E=17.439-25.589=-8.150$ \\
SB     & $\varepsilon_{2\Lambda}=8.483-2.399=6.083$ \\
       & $\varepsilon_{\Lambda \alpha}=8.774-11.595=-2.821$ \\
\hline
\end{tabular}
\end{center}
\end{table}

Summarizing this work, we have applied
the quark-model $YN$ and $YY$ interactions,
fss2 \cite{fss2,B8B8} and FSS \cite{FSS},
to the Faddeev calculation of the $\Lambda \Lambda \alpha$ system
for $\hbox{}^{\ 6}_{\Lambda \Lambda} \hbox{He}$,
in the new three-cluster Faddeev formalism
using two-cluster RGM kernels.
The $\Lambda \alpha$ $T$-matrix is generated
from the $\Lambda N$ effective force, which is derived
from the $\hbox{}^1S_0$ and $\hbox{}^3S_1$ $\Lambda N$ phase
shifts of fss2 by the supersymmetric inversion
method \cite{SB97}. With a single adjustable
parameter, this $\Lambda N$ force gives a realistic
description of the $\hbox{}^5_{\Lambda}\hbox{He}$ and
$\hbox{}^{9}_{\Lambda}\hbox{Be}$ systems \cite{BE9L}.
The $\Lambda \Lambda$ interaction of the quark-model
baryon-baryon interactions is therefore reliably examined
by solving the RGM $T$-matrix
in the $\Lambda \Lambda$-$\Xi N$-$\Sigma \Sigma$ coupled-channel
formalism, and by using it in the coupled-channel Faddeev
equation. Here we have used
only $\Lambda \Lambda \alpha$ configuration
and obtained $\Delta B_{\Lambda \Lambda}=1.41$ MeV for fss2,
as a measure of the the two-$\Lambda$ separation energy.
A simple Gaussian $\Lambda \Lambda$ potential, reproducing
the $\hbox{}^1S_0$ $\Lambda \Lambda$ phase shift of fss2,
yields $\Delta B_{\Lambda \Lambda}=1.91$ MeV.
Considering some repulsive effects from the Brueckner rearrangement
of the $\alpha$-cluster ($\sim$ 1 MeV) \cite{rearr}
and the quark Pauli principle between the $\alpha$ cluster
and the $\Lambda$ hyperon ($\sim$ 0.1 - 0.2 MeV) \cite{SU99},
we can conclude that the present results by fss2 are
in good agreement with the experimental value,
$\Delta B^{\rm exp}_{\Lambda \Lambda}=1.01 \pm 0.20$ MeV,
deduced from the Nagara event \cite{TA01}.
Together with previous several Faddeev calculations,
we have found that the model fss2 yields a realistic
description of many three-body systems,
including the three-nucleon bound state \cite{triton},
the hypertriton \cite{hypt},
$\hbox{}^9_{\Lambda}\hbox{Be}$ \cite{BE9L},
and $\hbox{}^{\ 6}_{\Lambda \Lambda} \hbox{He}$.  


\begin{acknowledgments}
This work was supported by Grants-in-Aid for Scientific
Research (C) from the Japan Society for the Promotion
of Science (JSPS) (Nos.~15540270, 15540284, and 15540292).
\end{acknowledgments}

\end{document}